\documentclass{aastex}
\usepackage{spr-astr-addons}
\usepackage{url}

\RequirePackage{color}

\newcommand{\teff}{T$_{\rm eff}$ }
\newcommand{\tsin}{T$_{\rm eff}$}

\begin{document}

\title{The solar, exoplanet and cosmological lithium problems}
\shorttitle{Short article title}
\shortauthors{Autors et al.}

\author{J. Mel\'endez}
\affil{Centro de Astrof\'{i}sica da Universidade do Porto, Rua das Estrelas, 4150-762 Porto, Portugal}
\and
\author{I. Ram\'{i}rez}\altaffilmark{2} \and \author{L. Casagrande} \and \author{M. Asplund}
\affil{Max Planck Institute for Astrophysics, Postfach 1317, 85741 Garching, Germany}
\and
\author{B. Gustafsson}
\affil{Institutionen f\"or fysik och astronomi, Uppsala universitet, Box 515, SE-75120 Uppsala, Sweden}
\and
\author{D. Yong}
\affil{Research School of Astronomy \& Astrophysics, Australian National University, Mount Stromlo Observatory, Cotter Road, Weston Creek, ACT 2611, Australia}
\and
\author{J. D. do Nascimento  Jr.} \and \author{M. Castro}
\affil{Departamento de F\'{i}sica Te\'orica e Experimental, Universidade Federal do Rio Grande do Norte, CEP: 59072-970 Natal, RN, Brazil}
\and
\author{M. Bazot}
\affil{Centro de Astrof\'{i}sica da Universidade do Porto}



\begin{abstract}
We review three Li problems. First, the Li problem in the Sun, for which some previous studies have argued that it may be Li-poor compared to other Suns. Second, we discuss the Li problem in planet hosting stars, which are claimed to be Li-poor when compared to field stars. Third, we discuss the cosmological Li problem, i.e. the discrepancy between the Li abundance in metal-poor stars (Spite plateau stars) and the predictions from standard Big Bang Nucleosynthesis. In all three cases we find that the
``problems'' are naturally explained by non-standard mixing in stars.
\end{abstract}



\section{The solar Li problem}

As illustrated in Fig. 1, the present day solar Li abundance
\citep[$A_{\rm Li}$\footnote{$A_X = \log (N_X/N_H) + 12$} = 1.05;][]{asp09} 
is much lower than the meteoritic Li abundance \citep[$A_{\rm Li}$ = 3.26;][]{asp09}. 
This large depletion in the observed solar Li abundance by a factor of 160 relative to the primordial solar system composition remains one of the most serious challenges of standard solar models, which, as illustrated in Fig. 1, destroy only a minor amount (0.06 dex) of Li \citep[e.g.][]{dan84}. It is important to note that modern models
using updated OPAL opacities predict too much lithium destruction 
during the pre-main sequence \citep[e.g.][]{pia02,dan03}, 
at variance with the observations of Li in solar-mass stars 
in young open clusters (see e.g. Fig. 6). When the new low solar abundances
\citep{asp09} are used the problem is reduced \citep{ses06},
but in general classical models have problems dealing with pre-main sequence convection,
so other parameters are invoked to reduce the efficient lithium destruction
during the pre-main sequence \citep[e.g.][]{ven98}.

\subsection{Comparison to solar analogs}

A comparison between the Sun and solar analogs of one solar mass and 
solar metallicity \citep{lam04} shows the Sun to be ``lithium-poor'' by a factor of 10. This apparent peculiarity in the solar Li abundance, led \cite{lam04} to suggest that the Sun may be of dubious value for calibrating non-standard models of Li depletion. However, Pasquini et al. (1994) have shown that there are solar-type stars that have a Li abundance as low as in the Sun. Nevertheless, \cite{pas94} comparison sample of solar-type stars span a wide range in stellar parameters (5400 K $<$ \teff $<$ 6100 K, 3.6 $<$ log $g$ $<$ 4.6, -1.6 $<$ [Fe/H] $<$ +0.2) and therefore they are not representative of one-solar-mass solar analogs. In Fig. 2, we restrict the comparison sample of \cite{pas94} to only solar analogs within $\pm$200K in solar effective temperature, $\pm$0.3 dex of the solar surface gravity and $\pm$0.3 dex of the solar metallicity. As we can see, these solar analogs seem to cluster in two groups, one with very high lithium abundances of $A_{\rm Li}$ $\sim$ 2.4, i.e., 20 times higher than solar, and the other group with Li abundances as low as solar. 
Why are there no solar analogs with intermediate Li abundances? 
Why the number of solar analogs with low Li abundances is much lower
than the number of analogs with high Li abundances? 
Could this be the reason why \cite{lam04} 
only found stars with high Li abundances around one solar mass? 
The lack of stars with intermediate Li abundances in the Pasquini et al. (1994) solar analog sample, and the lack of stars with both intermediate and low Li abundance around one solar mass in the \cite{lam04} sample are probably telling us that both samples have biases, perhaps due to a selection of stars mostly in one or two evolutionary stages, e.g., mainly young stars, which are known to have high Li abundances.

The recent work by \cite{pas08} for solar analogs and solar twins in the solar-age open cluster M67, shows that solar twins (M67 stars around solar effective temperature) have Li abundance as low as solar, but stars 100 or 200 K hotter span a broad range in Li abundances. So, it is important that the temperature scale of the comparison sample is accurate; otherwise offsets of about 100 K may introduce a bias in the
comparison between the Sun and stars.

\subsection{Comparison to solar twins}

Solar twins, stars with stellar parameters very similar to the Sun, are ideal targets to see if the Sun is normal (or not) in its Li abundance. Being so similar to the Sun, it is possible to obtain reliable stellar parameters, and provided the Sun and the twins are analyzed (and observed) in a consistent way, the temperature scale is accurate. 
Furthermore, being selected due to their similarity in colors and luminosity to the Sun, they should span a range of ages very close to solar, avoiding thus potential biases 
in the selection of comparison stars in only one evolutionary stage very different 
to the present Sun.

Solar twins have been searched for a long time, and although interesting 
solar twin candidates like 16 Cyg B (HD 186427)
were identified in the past, detailed analysis showed that
they were significantly different to the Sun \citep[see review by][]{cay96}.

When the first close solar twin (18 Sco) was found \citep{por97}, it seemed to have a Li abundance near solar, but much better data \citep[e.g.][]{mel06} showed that its Li abundance is actually three times higher than solar. One solar twin is certainly not an acceptable number for a comparison between the Sun and stars, so a large survey of solar twins was urgently needed. The two largest recent efforts for finding field solar twin stars are being undertaking by the group of Y. Takeda \citep[e.g.][]{tak07} and by our group 
\citep{mel06,mel07,mel09a,ram09}. Importantly, whenever possible, 
we are obtaining very high S/N for our sample stars, because otherwise 
only upper limits can be obtained for
their Li abundances. Indeed, as shown by \cite{tak07} in their Fig. 12, their data with S/N $\sim$150 can only estimate upper limits for stars with $A_{\rm Li}$ $<$ 1.5, i.e., they can only reliably determine Li abundances when they are three times higher than solar. 

Our solar twin survey has been performed mainly with the 2.7m telescope at McDonald observatory in the North and with the 6.5m Magellan Clay telescope at Las Campanas observatory in the South. We have also obtained some Keck+HIRES data in the North and VLT+UVES and HARPS data in the South. Our data has been taken at R = 60,000-110,000 and achieving S/N = 200-1000. The first pilot data set taken at Keck resulted in the
discovery of the second best solar twin, HD 98618 \citep{mel06}, about a decade after the discovery of the first solar twin 18 Sco. HD 98618 seems to be a solar twin as good as 18 Sco, and, as this twin, it has also a Li abundance three times higher than solar. Learning from the experience of our pilot Keck observations, we improved our criteria to select the best solar twins, empirically adjusting our \teff scale \citep{ram05} for
an apparent zero-point problem \citep{cas09}. 
This is probably the reason why our first solar twin run at McDonald was very successful. Besides confirming the solar twin nature of 18 Sco and HD 98618, we identified two additional solar twins, HIP 56948 and HIP 73815 \citep{mel07}, both with a low Li abundance similar to solar. HIP56948 remains to this date the star that most closely resembles the Sun, with a \teff similar to solar within 10 K, as recently confirmed by \cite{tak09} using Subaru+HDS observations. The year 2007 was very prolific for solar twin studies, besides the twins found by our group, \cite{tak07} reported the discovery of the fifth solar twin, HIP 110963. 
Interestingly, this solar twin has a high Li abundance of $A_{\rm Li}$ = 1.7. 
These five solar twins were starting to fill the Li desert seen in Fig. 2.

The year 2009 has been even better, with many more solar twins found using our McDonald data \citep{ram09} and Magellan+MIKE observations \citep{mel09a}, which together have found more than 30 stars very similar to the Sun. Thus, we are starting to have a large sample of solar twins for meaningful statistics, in particular for addressing the long-standing question on chemical peculiarities in the Sun \citep{mel09a,ram09,gus09}.

In Fig. 3 we show the Li abundance for solar twins and solar analogs 
that have been analyzed for Li in our survey. Most stars from the Magellan run have already been analyzed, but the McDonald Li analysis is just starting. Open circles show detections and filled circles upper limits. The Sun is also shown for comparison. The first important point to note in this plot is that, unlike the lack of stars with both intermediate and low Li in the \cite{lam04} sample, and the lack of stars with intermediate Li abundances in the \cite{pas94} sample, our sample nicely covers a broad range of Li abundances 0.6 $<$ $A_{\rm Li}$ $<$ 2.4, meaning probably that our sample is not affected by any significant selection bias.

In Fig. 4 we show the Li abundances vs. \teff of our solar twin and solar analog sample restricted to stars with mass within $\pm$3\% solar and [Fe/H] within $\pm$0.1 dex solar. The Sun does not look peculiar on this plot. One star with relatively high Li abundance stands out. As shown in Fig. 5, where Li is plotted as a function of age, the high Li abundance of this star is due to its young age. This plot has a smaller number of stars than Figs. 3-4 because here, besides the constraint to $\pm$3\% in mass and $\pm$0.1 dex in [Fe/H], we show only stars for which we could determine reliable ages ($\ge$ 2.5 sigma). This figure definitely shows that the solar Li abundance is not abnormal, at least not for a solar-metallicity solar-age one-solar-mass star, which at about 4.6 Gyr has already depleted a significant fraction of its original Li abundance.

A comparison of our solar twin results to one-solar-mass stars
in solar metallicity ($\pm$0.15 dex) open clusters
\citep[selected from the sample of][and including the 
results from Pasquini et al. 2008]{ses05} 
is shown in Fig. 6. Again, the agreement is excellent, and reinforces a strong correlation between Li depletion and age for one-solar-mass stars.

Non-standard models \citep[e.g.][]{mon00,chat05,xio09,don09} can reproduce the observed data, as shown in Figs. 7 and 8. We are in the process of obtaining better Li abundances and ages for our sample stars, which can potentially constrain the range of initial rotational velocities of our solar twins.

Based on the results shown above, we conclude that the solar Li
abundance is not peculiar but a product of depletion due to non-standard
mixing which affect both the Sun and the solar twins.

\section{Is Li depleted in stars with planets?}

Many works have compared stars with planets to
field stars instead of stars without detected planets.
This has been usually done to have a large comparison sample.

\cite{gon00} compared the Li abundance of stars with planets
to the Li abundance of field stars with detectable Li, finding that
(as previously believed for the Sun), 
stars with planets tend to have smaller Li abundances.
This apparent peculiarity in the comparison between stars with planets
and field stars could be due to a selection bias, as we have shown
in the previous section for the case of the Sun. Indeed, the same year,
\cite{rya00} showed that stars with planets have Li abundances 
indistinguishable from field stars, when stars of the
same temperature, age and composition were compared.

\cite{isr04} compared also planet hosting and field stars, showing
that indeed they have the same Li abundance, except perhaps for stars 
around solar \tsin. It is important to note that
this analysis is not fully consistent, as the comparison sample
used by Israelian et al. was taken from the literature \citep{che01}, 
potentially having problems with different temperature scales. Also,
as we have shown in the previous section, there are indeed many field
stars around solar temperature showing a low Li abundance, thus,
stars with planets probably do not have anomalous Li abundances
even around the solar \tsin.

\cite{tak05} also compared stars
with planets and field stars, showing that their Li abundances
are identical at any temperature except for a small narrow range
around 5850 K $\pm$ 50 K. This \teff range is so narrow that the
difference between stars with planets and field stars could
probably vanish when more comparison stars were included. Interestingly,
\cite{che06} found that the discrepancy could occur at
lower temperatures, around 5750 $\pm$ 50 K. Again, the lack
of Li-poor solar analogs, could be misleading
many authors to believe that stars with planets may be
different to field stars around solar \tsin, but our sample
of solar twins with low Li abundances weakens those claims.

In the largest homogeneous Li study of stars with planets and
field stars, \cite{luc06} concluded that there is no
discernible difference between planet hosts and comparison stars.

In Fig. 2 of \cite{gon08}, planet hosting stars also seem 
identical to field stars, except in the narrow \teff range of 5850 $\pm$ 50 K.
No conclusions can be made for stars with \teff $<$ 5800 K 
because only three planet hosts are available. Note that the
\cite{gon08} study is not homogeneous, as the Li abundances were obtained from
different literature works with probably different \teff scales
and different selection criteria. 
Thus, there could be potential selection
biases, as we have shown for the case of the apparent low solar Li 
abundance. Indeed, we have just finished the analysis of 4 planet
hosting stars and 6 stars without detected planets 
\citep{mel09a}, both around the above temperature range 
where \cite{gon08} found anomalously low Li abundances
in planet hosts.  As shown in Fig. 9, our planet hosting stars
do not show anomalously low Li abundances with respect to our
stars without detected planets.

We conclude that stars with and without planets probably
share similar Li abundances, at least to the level of
precision of current analyses, although a definitive
conclusion is not possible because Li depends on 
both mass and age (like we have shown for the Sun).
It is important that in future studies the comparison between 
planet hosts and stars without detected planets is performed 
for stars with similar parameters (mass, metallicity, age), and 
both samples must be analyzed in a consistent way. Any
claim for differences in Li between stars with and without planets
must be taken with skepticism unless reliable ages and
masses are determined for planet hosts and the comparison sample.

\section{The cosmological Li discrepancy}

One of the most important discoveries in the study of
the chemical composition of stars was made in 1982
by Monique and Fran\c cois Spite, who found an essentially constant
Li abundance in warm metal-poor stars (Spite \& Spite 1982), 
a result interpreted as a relic of
primordial nucleosynthesis. Due to its cosmological
significance, there have been many studies
devoted to Li in metal-poor field stars
\citep[e.g.][]{rya99,mel04,boe05,chap05,asp06,shi07,bon07,hos09,aok09},
with observed Li abundances at the lowest [Fe/H] 
from as low $A_{\rm Li}$  = 1.94 \citep{bon07}
to as high as $A_{\rm Li}$  = 2.37 \citep{mel04}.

Using the theory  of big bang nucleosynthesis (BBN) and
the baryon density obtained from WMAP data \citep{dun09}, 
a primordial Li abundance
of $A_{\rm Li}$  = 2.72$_{-0.06}^{+0.05}$ is predicted 
\citep[][see also e.g. \cite{ste07}]{cyb08}, which is
a factor of 2-6 times higher than the Li abundance inferred
from halo stars.
There have been many theoretical studies on non-standard
BBN trying to explain the cosmological Li discrepancy by
exploring the frontiers of new physics \citep[e.g.][]{coc09,jed09,koh09}. 
Alternatively, the Li problem 
could be explained by a reduction of the original Li stellar abundance
due to internal processes (i.e., by stellar depletion).
In particular, stellar models including atomic diffusion and mixing can 
deplete a significant fraction of the initial Li content \citep{ric05,pia08},
although such models depend on largely unconstrained free parameters.
Due to the uncertainties in the Li abundances
and to the limited samples available, only limited comparisons of 
models of Li depletion with stars in a broad range of mass and
metallicities  have been performed.

In order to provide meaningful comparisons with stellar depletion
models, precise Li abundances for a large sample of stars  are needed. 
We have recently finished such a study \citep{mel09b}, 
achieving errors in
Li abundance lower than 0.035 dex, for a large sample 
of metal-poor stars (-3.5 $<$ [Fe/H] $<$ -1.0), for
the first time with precisely determined masses 
in a relatively broad mass range (0.6-0.9 M$_\odot$).

Our recent work shows that Li is depleted in Spite plateau stars.
As can be seen in Fig. 10, the spread of the Spite plateau at
any metallicity is much larger than the error bar. Also, there
seems to be a correlation with \teff at any probed metallicity.
Actually the correlation is better when Li is plotted versus
stellar mass (Fig. 11), showing thus that Li has been depleted
in Spite plateau stars at any metallicity. 
In this figure we confront the 
stellar evolution predictions of \cite{ric05} with our inferred
stellar masses and Li abundances. 
The models include the effects of atomic diffusion, 
radiative acceleration and gravitational settling
but moderated by a parametrized turbulent mixing;
so far only the predictions for [Fe/H]=$-$2.3 are available
for different turbulent mixing models.  
The agreement is very good when adopting a turbulent model of T6.25 
(see Richard et al. for the meaning of this notation)  
and an initial $A_{\rm Li}  = 2.64$. The stellar Li abundances used 
above were obtained with the latest MARCS models \citep{gus08}, 
but if we use instead the Kurucz convective overshooting
models, then the required initial abundance to explain our observational data 
would correspond to $A_{\rm Li} = 2.72$.

Our results imply that the Li abundances observed in Li plateau
stars have been depleted from their original values and therefore
do not represent the primordial Li abundance.
It appears that the observed Li abundances in metal-poor stars
can be reasonably well reconciled with the predictions 
from standard Big Bang nucleosynthesis \citep[e.g.][]{cyb08}
by means of more realistic stellar evolution models that
include Li depletion through diffusion and turbulent mixing
\citep{ric05}. 
We caution however, that, although encouraging, our results 
should not be viewed as proof of the correctness of the 
diffusion+turbulence models of Richard et al. models until the
free parameters required for the stellar modeling are 
better understood from basic physical principles.
In this context, new physics should not be discarded yet
as a solution of the cosmological Li discrepancy, 
as perhaps the low Li-7 abundances in metal-poor stars 
might be a signature of supersymmetric particles in the early universe,
which could also explain the Li-6 detections claimed for
some metal-poor stars \citep{asp06,asp08}.

\acknowledgments
This work has been partially supported by FCT (project
PTDC/CTE-AST/65971/2006, and Ciencia 2007 program)
and by the FCT/CAPES cooperation agreement between
Portugal and Brazil.
J.M. thanks ANSTO for travel support.

\nocite{*}


\begin{figure}[t]
\plotone{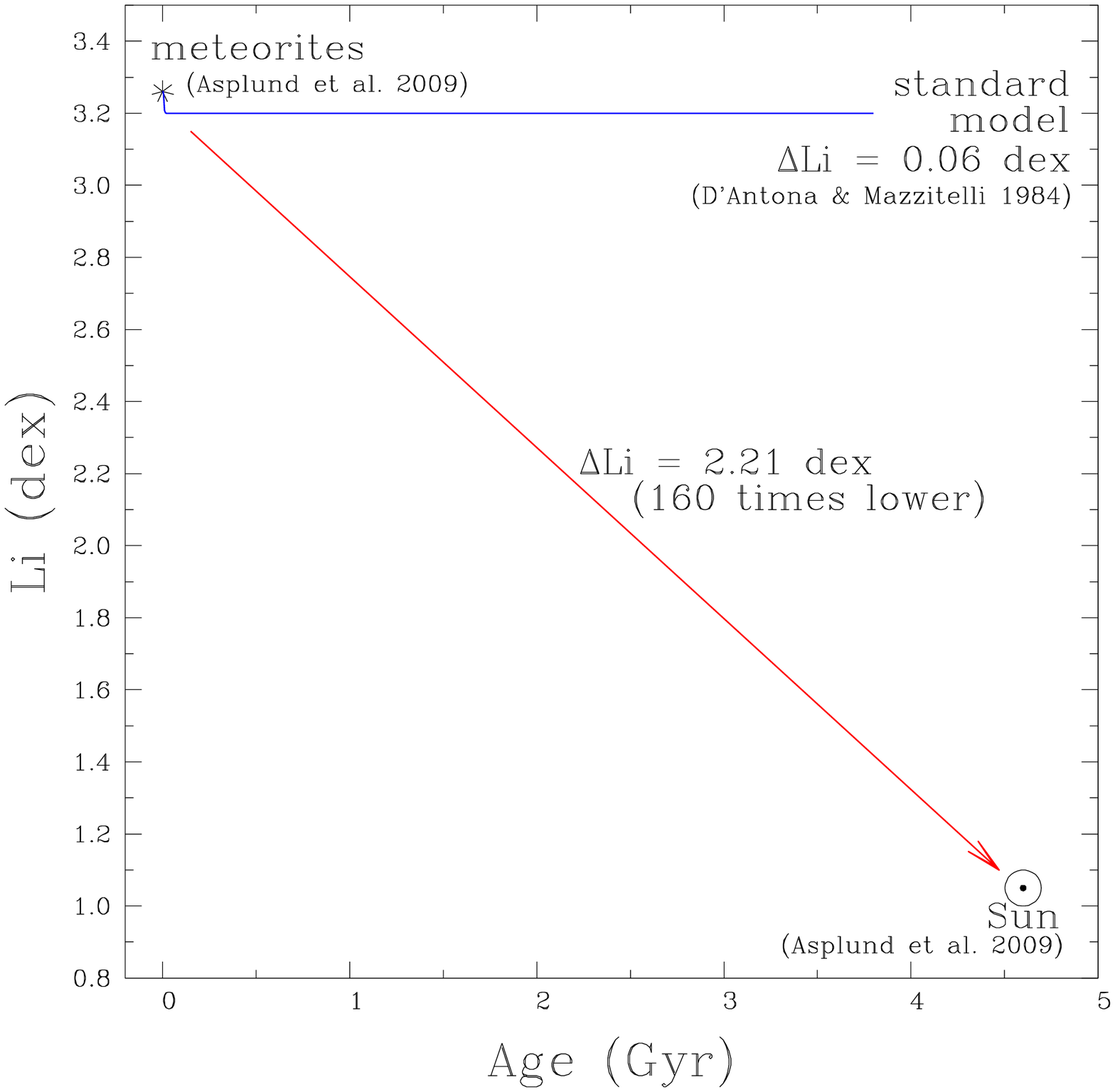}
\caption{The standard solar model \citep[e.g.][]{dan84} 
can not reproduce the large
decrease in Li abundance at the present solar age.
The Sun is represented by the $\odot$ symbol
} 
\end{figure}

\begin{figure}[t]
\plotone{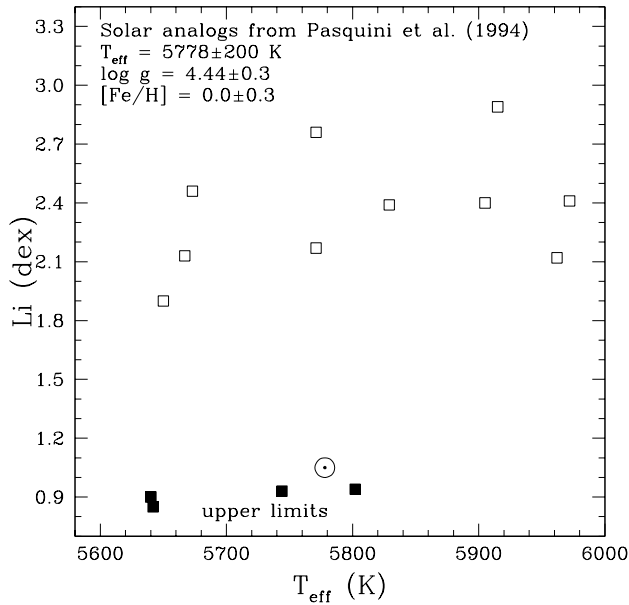}
\caption{%
Li in the solar analog sample of \cite{pas94} restricted to 
stars with \teff within 200 K solar,
and both log $g$ and [Fe/H] within 0.3 dex solar. Open symbols are detections, filled
symbols are upper limits} 
\end{figure}

\begin{figure}[t]
\plotone{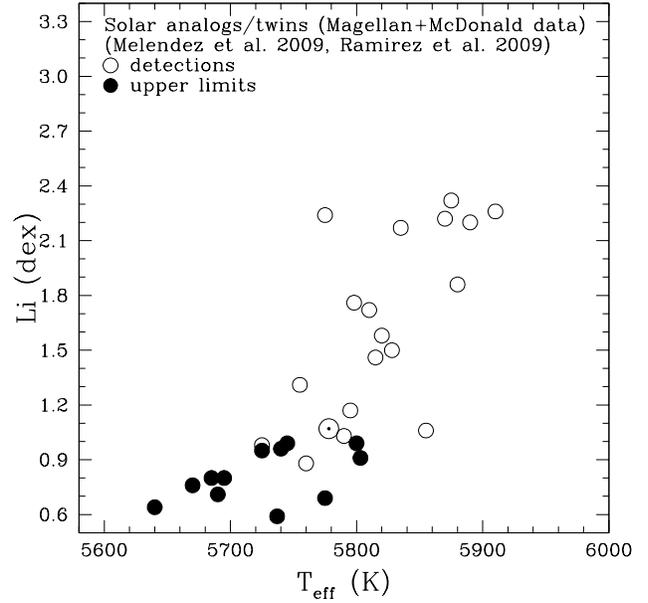}
\caption{%
Magellan and McDonald solar analogs and twins
covering a relatively broad range in stellar parameters
(5640 K$<$ \teff $<$ 5920 K), 4.31 $<$ log $g$ $<$ 4.56, 
and $-0.45 <$ [Fe/H] $< +0.45$). Open circles
are detections and filled circles are upper limits
} 
\end{figure}

\begin{figure}[t]
\plotone{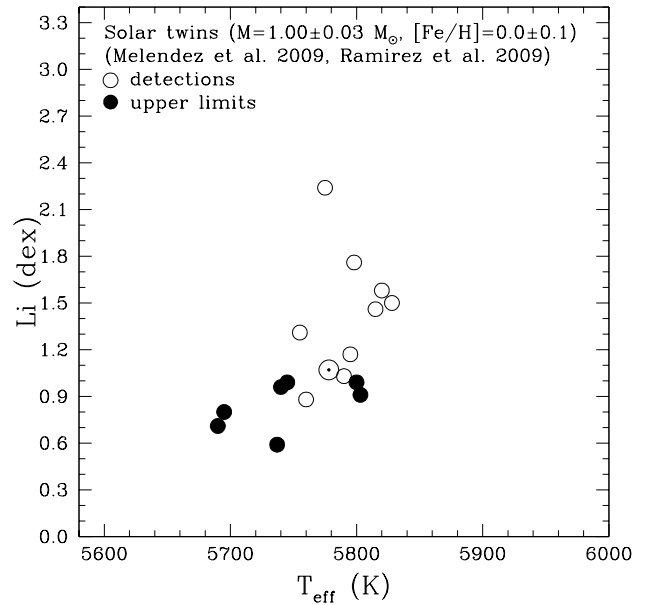}
\caption{%
Magellan and McDonald solar twins
with one solar mass ($\pm$ 3\%) and solar
[Fe/H] ($\pm$ 0.1 dex). Symbols as in Fig. 3
} 
\end{figure}

\begin{figure}[t]
\plotone{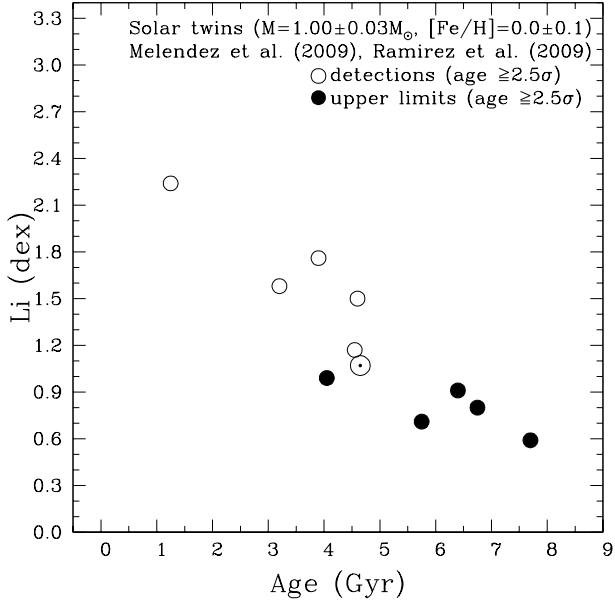}
\caption{%
Li as a function of age for our 
Magellan and McDonald solar twins
with one solar mass ($\pm$ 3\%) and solar
[Fe/H] ($\pm$ 0.1 dex). Symbols as in Fig. 3.
Only stars with reliable ages ($\ge$ 2.5 $\sigma$) are shown
} 
\end{figure}

\begin{figure}[t]
\plotone{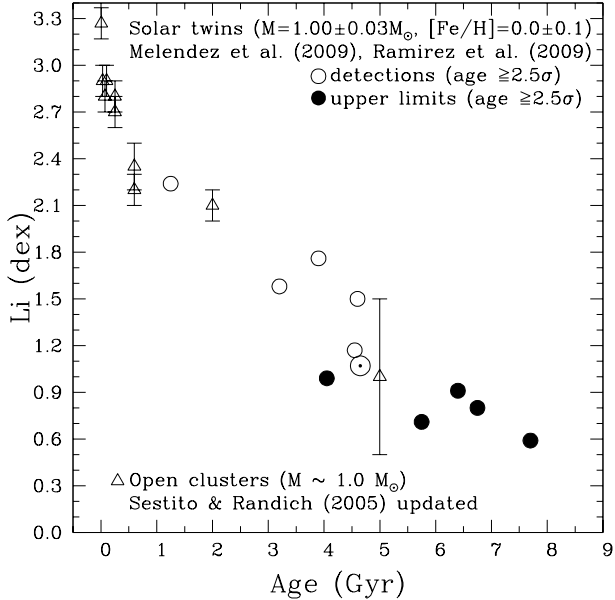}
\caption{%
Li for our solar twins
with one solar mass ($\pm$ 3\%) and solar
[Fe/H] ($\pm$ 0.1 dex), and for one-solar-mass
stars in solar metallicity ($\pm$0.15 dex) 
open clusters selected from
\cite{ses05}, although for M67 we used the
sample of \cite{pas08}. Field stars are shown
as circles while open clusters (NGC 2264, IC2602/IC2391, Pleiades,
Blanco 1, NGC6475, M34, Coma Berenices, Hyades, NGC762, M67) with triangles}
\end{figure}

\begin{figure}[t]
\plotone{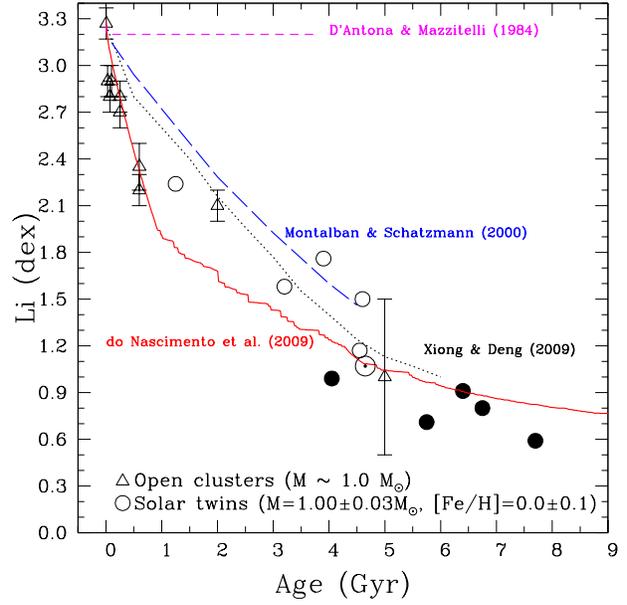}
\caption{%
Comparison of non-standard solar models to field
and open cluster stars}
\end{figure}

\begin{figure}[t]
\plotone{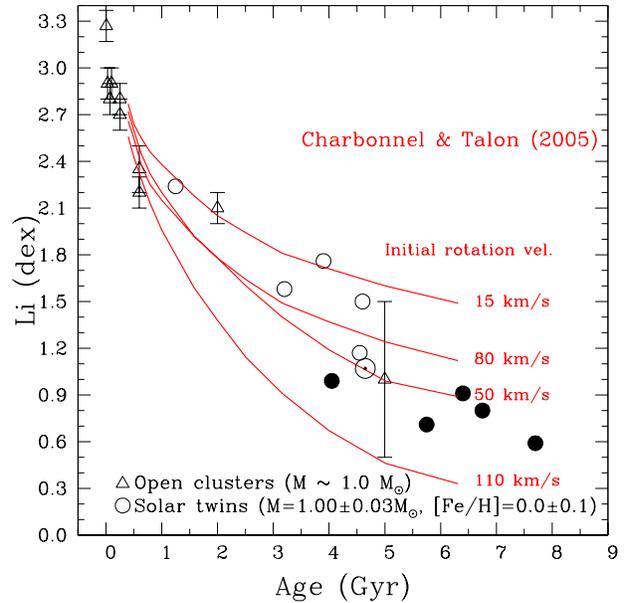}
\caption{%
Non-standard models of Li depletion by
\cite{chat05} for different initial rotation velocities}
\end{figure}

\begin{figure}[t]
\plotone{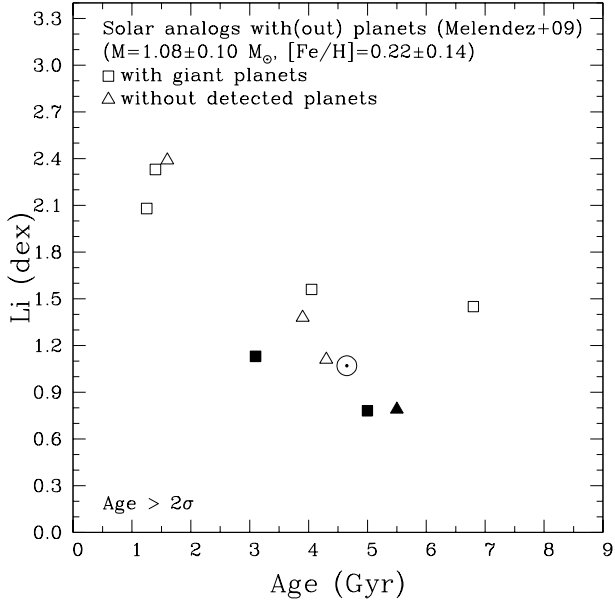}
\caption{%
Li in stars with (squares) and without (triangles)
detected planets from the solar analog sample
of \cite{mel09a}. Open symbols are detections and
filled symbols are upper limits}
\end{figure}

\begin{figure}[t]
\plotone{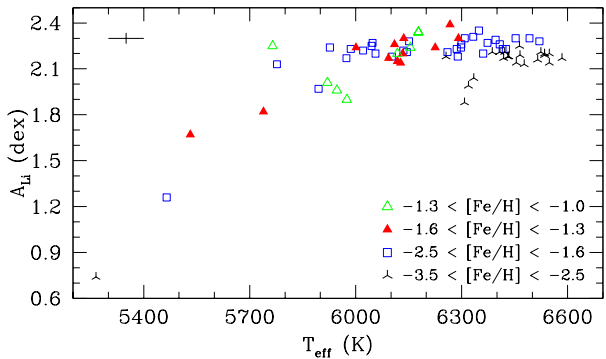}
\caption{%
Li abundances vs. \teff for our sample of metal-poor stars in different metallicity
ranges. The spread at any given metallicity is much larger than the
error bar. Figure taken from \cite{mel09b}}
\end{figure}

\begin{figure}[t]
\plotone{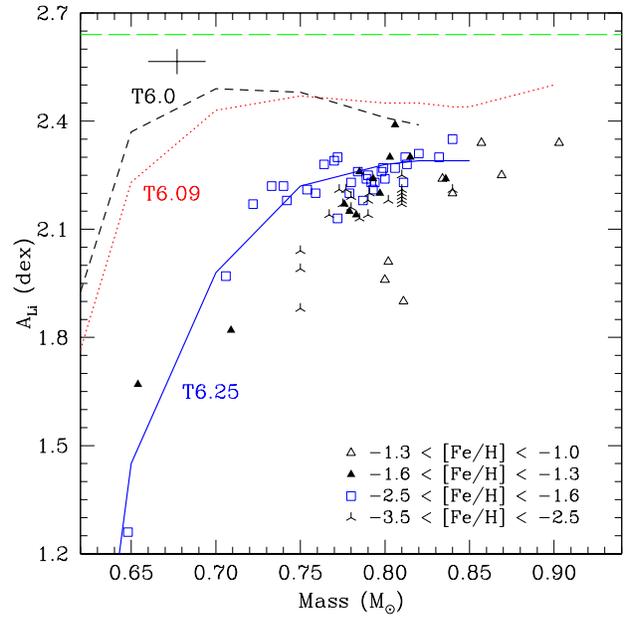}
\caption{%
Li abundances as a function of stellar mass in different
metallicity ranges. Models at [Fe/H] = $-2.3$ including diffusion and
T6.0 (short dashed line), T6.09 (dotted line) 
and T6.25 (solid line) turbulence \citep{ric05} are shown. 
The models have been rescaled to an initial $A_{\rm Li}$=2.64 (long dashed line).
Figure taken from \cite{mel09b}}
\end{figure}

\end{document}